\documentclass[twocolumn]{aastex63}

\usepackage{amsmath,bm}

\usepackage{ulem}
\renewcommand{\sout}{\bgroup \color[rgb]{1,0,0}\ULdepth=-.5ex \ULset}

\received{August 1, 2019}
\revised{\today}
\accepted{}

\begin{document}

\title{Ruling out the supersoft high-density symmetry energy
from the discovery of PSR J0740+6620 with mass $2.14^{+0.10}_{-0.09}M_\odot$}

\author{Ying Zhou}
\affiliation{School of Physics and Astronomy and Shanghai Key Laboratory for
Particle Physics and Cosmology, Shanghai Jiao Tong University, Shanghai 200240, China}

\author{Lie-Wen Chen}
\affiliation{School of Physics and Astronomy and Shanghai Key Laboratory for
Particle Physics and Cosmology, Shanghai Jiao Tong University, Shanghai 200240, China}

\correspondingauthor{Lie-Wen Chen}
\email{lwchen$@$sjtu.edu.cn}



\begin{abstract}

Using the very recently reported mass $2.14^{+0.10}_{-0.09}M_\odot$ of PSR J0740+6620
together with the data of finite nuclei and the constraints on the equation of state of symmetric nuclear matter
at suprasaturation densities from flow data in heavy-ion collisions, we show that
the symmetry energy $E_{\rm sym}(n)$ cannot be supersoft so that it becomes negative at suprasaturation
densities in neutron stars (NSs) and thus may make the NS have a pure neutron matter core.
This is in contrast to the fact that using the mass $2.01 \pm 0.04 M_\odot$ of PSR J0348+0432 as the NS maximum mass
cannot rule out the supersoft high-density $E_{\rm sym}(n)$.
Furthermore, we find the stiffer high-density $E_{\rm sym}(n)$ based on the existence
of $2.14M_\odot$ NSs leads to a strong constraint of $\Lambda_{1.4} \ge 348^{+88}_{-51}$
for the dimensionless tidal deformability of the canonical $1.4M_\odot$ NS.

\end{abstract}



\section{Introduction} \label{sec:intro}

The density dependence of nuclear symmetry energy $E_{\rm sym}(n)$,
which characterizes the isospin dependence of the equation of state (EOS) of nuclear matter,
is fundamentally important due to its
multifaceted roles in nuclear physics and
astrophysics~\citep{Dan02,Lat04,Ste05,Bar05,LCK08,Bal16,Oze16,Lat16,Wat16,Oer17,Wol18,Bla18,LiBA19}.
Theoretically, it is still a big challenge to calculate the $E_{\rm sym}(n)$ directly from
the first-principle non-perturbative QCD~\citep{Bra14}, and
currently information on the $E_{\rm sym}(n)$ is mainly obtained
in the effective models.
So far essentially all available nuclear effective models
have been used to calculate
the $E_{\rm sym}(n)$, and the results can be roughly classified equally
into two groups (see, e.g., Refs.~\citep{Sto03,ChenLW17}), i.e., a group where
the $E_{\rm sym}(n)$ increases with the density $n$, and the other group where
the $E_{\rm sym}(n)$ first increases with $n$ and then decreases above
a certain suprasaturation density and even becomes negative at
high densities.
The $E_{\rm sym}(n)$ in the latter group is generally regarded as soft,
and here we regard the $E_{\rm sym}(n)$ as {\it supersoft} if it becomes negative at
the suprasaturation densities inside neutron stars (NSs).
In this sense, the supersoft $E_{\rm sym}(n)$ may make the NS have
a pure neutron matter (PNM) core, which will have important implications on
the chemical composition and cooling mechanisms of protoneutron stars~\citep{Lat91,Sum94,Pra97},
the critical densities for the appearance of hyperons~\citep{Pro19} and antikaon condensates~\citep{Lee96,Kub99} in NSs,
the NS mass-radius relations~\citep{Pra88,Eng94}, and
the possibility of a mixed quark-hadron phase~\citep{Kut00,Wu19} in NSs.

Unfortunately, the high-density behavior of the $E_{\rm sym}(n)$ is still very elusive, although
the $E_{\rm sym}(n)$ at subsaturation densities has been relatively well
determined from analyzing the data of finite nuclei (see, e.g., Refs.~\citep{Zha13,Bro13,Dan14,Zha15}.
In terrestrial laboratories,
the high-density nuclear matter can be produced only in high-energy
heavy-ion collisions, and presently the resulting high-density
$E_{\rm sym}(n)$ from analyzing the data in heavy-ion collisions can be either supersoft or stiff, strongly depending on the
models and data~\citep{Xiao09,Fen10,Rus11,Xie13,Coz13,Hon14,Rus16,Zha17}.
In nature,
the NSs provide an ideal astrophysical site to explore the high-density $E_{\rm sym}(n)$.
In particular, the dimensionless tidal deformability $\Lambda_{M}$ for a NS with mass $M$,
which is specifically sensitive to the NS radius and thus the high-density
$E_{\rm sym}(n)$,
can be extracted from the gravitational wave (GW) signal of
the binary neutron star (BNS) merger~\citep{Hin08,Fla08,Hin10,Vin11,Dam12}.
Actually,
the limit of $\Lambda_{1.4} \le 580$ from the recent GW signal
GW170817~\citep{Abb17NSMerger,LAbb17EM}
already can exclude too stiff high-density $E_{\rm sym}(n)$~\citep{ZhouY19}.
In addition,
the existence of large mass NSs may set a lower limit for the
high-density $E_{\rm sym}(n)$, but we note that
the observed mass $2.01 \pm 0.04 M_\odot$ of PSR J0348+0432~\citep{Ant13} is still consistent with
the supersoft high-density $E_{\rm sym}(n)$~\citep{ZhouY19}.

Very recently, a millisecond pulsar J0740+6620 with mass
$2.14^{+0.10}_{-0.09}M_\odot$ ($68.3\%$ credibility interval) was
reported~\citep{Cro19Mmax} by combining the relativistic Shapiro
delay data taken over 12.5 years at the North American Nanohertz Observatory for
Gravitational Waves with recent orbital-phase-specific observations
using the Green Bank Telescope. This
pulsar may hence replace the previously reported heaviest
PSR J0348+0432 with mass $2.01 \pm 0.04 M_\odot$~\citep{Ant13}
and set a new record for the maximum mass of NSs. It is thus
interesting to examine whether this new heaviest NS
can give new insight on the high-density $E_{\rm sym}(n)$.

In this work, using the data of finite nuclei together with the constraints
on the EOS of symmetric nuclear matter (SNM) at suprasaturation densities from heavy-ion collisions,
we show the existence of NSs with mass $2.14M_\odot$ can rule out the supersoft
$E_{\rm sym}(n)$,
although the largest NS mass $2.01M_\odot$ cannot.
We further find the stiffer lower limit of the high-density $E_{\rm sym}(n)$
from the existence of NSs with mass $2.14M_\odot$
leads to a quite large lower bound value for $\Lambda_{1.4}$, 
i.e., $\Lambda_{1.4} \ge 348^{+88}_{-51}$.

\section{Model and method}

\subsection{Nuclear matter EOS}

For an isospin asymmetric nuclear matter with neutron (proton) number density
$n_{\rm n}$ ($n_{\rm p}$), its EOS $E(n, \delta)$ is usually
expressed as the binding energy per nucleon as a function of the nucleon number density
$n=n_{\rm n}+n_{\rm p}$ and the isospin asymmetry $\delta=(n_{\rm n}-n_{\rm p})/n$.
The $E(n, \delta)$ can be expanded in terms of $\delta$ as
\begin{equation}\label{EOS}
E(n,\delta)=E_0(n)+E_{\rm sym}(n)\delta^2+\cdots,
\end{equation}
where $E_0(n)=E(n,\delta=0)$ is the EOS of SNM,
and the symmetry energy $E_{\rm sym}(n)$ is defined by
\begin{eqnarray}\label{Eq.Esym}
E_{\rm sym}(n)=\left.\frac{1}{2!}\frac{\partial^{2}
    E(n,\delta)}{\partial\delta^2}\right|_{\delta=0}.
\end{eqnarray}
It should be mentioned that the odd-order terms of $\delta$ vanish
in Eq. (\ref{EOS}) due to the exchange symmetry between protons and neutrons
in nuclear matter.
At the saturation density $n_0$,
the $E_0(n)$ can be expanded in $\chi=(n-n_0)/3n_0$ as
\begin{equation}\label{Eq.E0}
  E_0(n)=E_0(n_0)+\frac{K_0}{2!}\chi^2+\frac{J_0}{3!}\chi^3+\cdots,
\end{equation}
where $E_0(n_0)$ is the binding energy per nucleon of SNM at $n_0$,
$K_0 =  \left. 9 n_0^2\frac{d^2 E_0(n)}{dn^2}\right|_{n=n_0}$ is
the incompressibility coefficient, and
$J_0 =  \left. 27 n_0^3\frac{d^3 E_0(n)}{dn^3}\right|_{n=n_0}$
is the skewness coefficient.

Around a reference density $n_r$, the $E_{\rm sym}(n)$ can be expanded
in $\chi_r=(n-n_r)/3n_r$ as
\begin{eqnarray}\label{Eq.Esym0}
E_{\rm sym}(n) & = & E_{\rm sym}(n_r) + L(n_r)\chi_r + \frac{K_{\rm sym}(n_r)}{2!}\chi_r^2+\cdots,
\end{eqnarray}
where $L (n_r) = \left. 3n_r\frac{dE_{\rm sym}(n)}{dn}\right|_{n=n_r}$ is
the density slope parameter and
$K_{\rm sym}(n_r) = \left. 9n_r^2\frac{d^2 E_{\rm sym}(n)}{dn^2}\right|_{n=n_r}$
is the density curvature parameter.
At $n_r=n_0$, the $L(n_r)$ and $K_{\rm sym}(n_r)$ are reduced, respectively,
to the well-known $L \equiv L (n_0)$ and $K_{\rm sym} \equiv K_{\rm sym}(n_0)$, which characterize
the density dependence of the $E_{\rm sym}(n)$ around $n_0$.

\subsection{The extended Skyrme-Hartree-Fock model}
In this work, we use a single theoretical model, namely,
the extended Skyrme-Hartree-Fock (eSHF) model~\citep{Cha09,Zha16} to
simultaneously describe nuclear matter, finite nuclei
and neutron stars.
Compared to the standard SHF model (see, e.g., Ref.~\citep{Cha97}),
the eSHF model contains additional momentum- and density-dependent two-body forces
to effectively simulate the momentum dependence of the three-body forces
and can describe very well the
properties of nuclear matter, finite nuclei
and neutron stars~\citep{Zha16}, which involve a wide
density region from subsaturation to suprasaturation densities.
We would like to emphasize that the density dependence of nuclear matter EOS and the
$E_{\rm sym}(n)$ from the eSHF model is very flexible.
In particular, within the eSHF model, the high-density $E_{\rm sym}(n)$ could be positive or negative
while the $E_{\rm sym}(n)$ at saturation and subsaturation densities
can be in nice agreement with the nuclear experimental constraints~\citep{Zha16}.
Accordingly, the eSHF model is especially suitable
for our present motivation to explore the possibility for the existence of the
supersoft high-density $E_{\rm sym}(n)$.

The extended Skyrme effective nucleon-nucleon interaction
is taken to have a zero-range,
density- and momentum-dependent form~\citep{Cha09,Zha16}, i.e.,
\begin{eqnarray}\label{Eq:SHF}
v(\bm{r}_i,\bm{r}_j)&=& t_0(1+x_0 P_\sigma)\delta(\bm{r})
    +\frac{1}{6}t_3(1+x_3 P_\sigma)
    n^\alpha(\bm{R}) \delta(\bm{r}) \nonumber \\
&+& \frac{1}{2}t_1(1+x_1 P_\sigma)[K'^2\delta(\bm{r})
    +\delta(\bm{r})K^2] \nonumber\\
&+&t_2(1+x_2 P_\sigma)\bm{K}'\cdot
    \delta(\bm{r}){\bm{K}} \nonumber\\
&+&\frac{1}{2}t_4(1+x_4P_\sigma)
    [K'^2 \delta(\bm{r}) n(\bm{R})
    +n(\bm{R})\delta(\bm{r})  K^2] \nonumber\\
&+& t_5(1+x_5 P_\sigma)\bm{K}'\cdot
    n(\bm{R}) \delta(\bm{r}){\bm{K}} \nonumber \\
&+&iW_0(\bm{\sigma}_i+\bm{\sigma}_j)\cdot
    [\bm{K}'\times\delta(\bm{r}){\bm{K}}],
\end{eqnarray}
where we have $\bm{R}=(\bm{r}_i+\bm{r}_j)/2$
and $\bm{r}=\bm{r}_i-\bm{r}_j$, $P_\sigma=(1+\bm{\sigma}_i\cdot\bm{\sigma}_j)/2$
is the spin exchange operator,
and $\bm{\sigma}_i$ ($\bm{\sigma}_j$) is the Pauli spin matrix.
In addition, the relative momenta operators
${\bm{K}}=(\bm{\nabla}_i-\bm{\nabla}_j)/2i$
and $\bm{K}'=-(\bm{\nabla}_i-\bm{\nabla}_j)/2i$
act on the right and left of the wave function, respectively.
The interaction includes $14$ independent model parameters,
i.e., the $13$ Skyrme force parameters $\alpha$,
$t_0\sim t_5$, $x_0\sim x_5$,
and the spin-orbit coupling constant $W_0$.
The $13$ Skyrme force parameters can be expressed explicitly in terms of
the following $13$ macroscopic quantities (pseudo-parameters)~\citep{Zha16}:
$n_0$, $E_{0}(n_0)$, $K_0$, $J_0$,
$E_{\rm sym}(n_r)$, $L(n_r)$, $K_{\rm sym}(n_r)$,
the isoscalar effective mass $m_{s,0}^{\ast}$,
the isovector effective mass $m_{v,0}^{\ast}$,
the gradient coefficient $G_S$,
and the symmetry-gradient coefficient $G_V$,
the cross gradient coefficient $G_{SV}$,
and the Landau parameter $G_0'$ of SNM in the spin-isospin channel.
For the motivation of the present work, instead of directly using the
$13$ Skyrme force parameters,
it is very convenient to use the
$13$ macroscopic quantities in the eSHF calculations for nuclear matter,
finite nuclei and neutron stars, and the details can be found in
Ref.~\citep{Zha16}.

\subsection{Tidal deformability of neutron stars}
The tidal deformability (polarizability) $\lambda$ of NSs can be thought of as
the NS fundamental \textit{f}-modes
with spherical harmonic index $l=2$ which
can be treated as forced and damped harmonic oscillators
driven by the external tidal field of the NS's companion.
The $\lambda$ is defined as the oscillation response coefficient
\citep{Fla08}, namely, the ratio of the neutron star's quadrupole moment
$ Q_{ij} $ to the companion's perturbing tidal field
${\cal E}_{ij}$ (in units with $c=G=1$ in this work)~\citep{Fla08,Hin08},
i.e., $\lambda = -Q_{ij} /{\cal E}_{ij}$.
The $\lambda$ is related to the dimensionless quadrupole
tidal Love number $ k_2 $ and  the NS radius $R$ by the relation
$\lambda=\frac{2}{3}k_2 R^5$.
For a NS with mass $M$, the dimensionless tidal deformability $\Lambda_{M}$ is
conventionally defined as
\begin{eqnarray}
\Lambda_{M}=\frac{2}{3}k_2(R/M)^5.
\end{eqnarray}
The Love number $k_2$ depends on the details of the NS structure
and it can be evaluated by~\citep{Hin08}
\begin{eqnarray}
  k_2 & = & 1.6C^5(1-2C)^2 [2-y+2C(y-1)] \nonumber\\
  &\times& \{ 2C [6-3y+3C(5y-8)] \nonumber\\
  &   +  & 4C^3 [ 13-11y+C(3y-2)+2C^2(1-y) ] \nonumber\\
  &   +  & 3(1-2C)^2 \ln{(1-2C)[2-y+2C(y-1)]}\}^{-1},~~~~
\end{eqnarray}
where $C=M/R$ is the NS compactness
and $y=y(R)$ is determined by solving the following
first-order differential equation:
\begin{equation}\label{dydr}
  \frac{dy(r)}{dr}=-\frac{y(r)^2+y(r)F(r)+r^2Q(r)}{r},
\end{equation}
with
\begin{eqnarray}
  F(r)&= &\frac{r-4\pi r^3[{\cal E}(r)-P(r)]}{r-2M(r)},\\
  Q(r)&= &\frac{4\pi r\left[5{\cal E}(r)+9P(r)
        +\frac{{\cal E}(r)+P(r)}{C_s^2}
        -\frac{6}{4\pi r^2}\right]}{r-2M(r)} \nonumber\\
      &  &-4\left\lbrace\frac{M(r)+4\pi r^3 P(r)}{r[r-2M(r)]}\right\rbrace^2.
\end{eqnarray}
In the above, $C_s^{2} \equiv dP(r)/d{\cal E}(r)$ is the squared sound speed of the NS matter.
Eq.~(\ref{dydr}) for dimensionless $ y(r) $ must be integrated with
the general relativistic equations of
hydrostatic equilibrium, namely, the famous Tolman-Oppenheimer-Volkoff (TOV)
equations \citep{Tol39,Opp39}:
\begin{eqnarray}\label{TOV}
\frac{dP(r)}{dr}&=&-\frac{[{\cal E}(r)+P(r)]
  [M(r)+4\pi r^3 P(r)]}{r [r-2M(r)]}, \\
  \frac{dM(r)}{dr}&=&4\pi r^2{\cal E}(r),
\end{eqnarray}
where $r$ is the radial coordinate,
$M(r)$ is the enclosed mass inside the radius $r$, and
${\cal E}(r)$ ($P(r)$) is the energy density (pressure) at $r$.
The boundary condition for Eq. (\ref{dydr}) is $y(0) =2$ \citep{Pos10}.
For a given NS matter EOS $P({\cal E})$,
one can calculate the NS mass $M$, radius $R$,
Love number $k_2$, and $\Lambda_M$ with
various central densities for the NS.

The NS contains core, inner crust, and outer crust.
The density $n_{\rm{out}}$ separating the inner and outer crusts
is taken to be $2.46 \times 10^{-4}~\rm{fm}^{-3}$,
and the core-crust transition density $n_{\rm t}$ is evaluated self-consistently
by a dynamical approach~\citep{XuJ09}.
We assume here that the core is composed of $\beta$-stable and electrically neutral
$npe\mu$ matter and its EOS can be calculated
within the eSHF model.
For the inner crust between densities $n_{\rm out}$ and $n_{\rm t}$,
the EOS is constructed by interpolating with
$P=a+b{\cal E}^{4/3}$ due to its complicated structure~\citep{Car03}.
For the outer crust, we employ the well-known Baym-Pethick-Sutherland EOS in the density region of
$6.93 \times 10^{-13}~{\rm fm}^{-3}<n<n_{\rm out}$ and Feynman-Metropolis-Teller EOS
for $n<6.93\times10^{-13}~{\rm fm}^{-3}$~\citep{Bay71,Iid97}.
The causality condition $dP/d{\cal E}\le 1$ is guaranteed for all the NS calculations in the present work.

\begin{table}[tbp]
\centering
\caption{Experimental data on the binding energies $E_{\rm{B}}$
(12 spherical even-even nuclei)~\citep{Wan17},
the charge r.m.s. radii $r_{\rm{c}}$ (9 nuclei)~\citep{Ang13,Fri95,LeBla05},
the isoscalar giant monopole resonance (GMR) energies $E_{\rm{GMR}}$
and its experimental error (4 nuclei)~\citep{You99},
and 5 spin-orbit energy level splittings $\epsilon_{\rm{ls}}^{A}$~\citep{Vau72}.
Here $\nu(\pi)$ denotes neutron(proton).
\label{expdata}}
\begin{tabular}{ccccc}
\tableline
\tableline
$^{A}$X       & $E_{\rm B}$(MeV) & $r_{\rm c}$(fm) & $E_{\rm GMR}$(MeV) & $\epsilon_{\rm ls}^{A}$(MeV) \\
\tableline
${}^{16}$O    & $-127.619$       & 2.6991          &                    & 6.10(1p$\nu$) \\
              &                  &                 &                    & 6.30(1p$\pi$) \\
${}^{40}$Ca   & $-342.052$       & 3.4776          &                    &               \\
${}^{48}$Ca   & $-416.001$       & 3.4771          &                    &               \\
${}^{56}$Ni   & $-483.995$       & 3.7760          &                    &               \\
${}^{68}$Ni   & $-590.408$       &                 &                    &               \\
${}^{88}$Sr   & $-768.468$       & 4.2240          &                    &               \\
${}^{90}$Zr   & $-783.898$       & 4.2694          & 17.81$\pm$0.35     &               \\
${}^{100}$Sn  & $-825.300$       &                 &                    &               \\
${}^{116}$Sn  & $-988.681$       & 4.6250          & 15.90$\pm$0.07     &               \\
${}^{132}$Sn  & $-1102.84$       &                 &                    &               \\
${}^{144}$Sm  & $-1195.73$       & 4.9524          & 15.25$\pm$0.11     &               \\
${}^{208}$Pb  & $-1636.43$       & 5.5012          & 14.18$\pm$0.11     & 1.32(2d$\pi$) \\
              &                  &                 &                    & 0.89(3p$\nu$) \\
              &                  &                 &                    & 1.77(2f$\nu$) \\
\tableline
\end{tabular}
\end{table}

\begin{figure*}[tbp]
\centering
\includegraphics[width=0.8\linewidth]{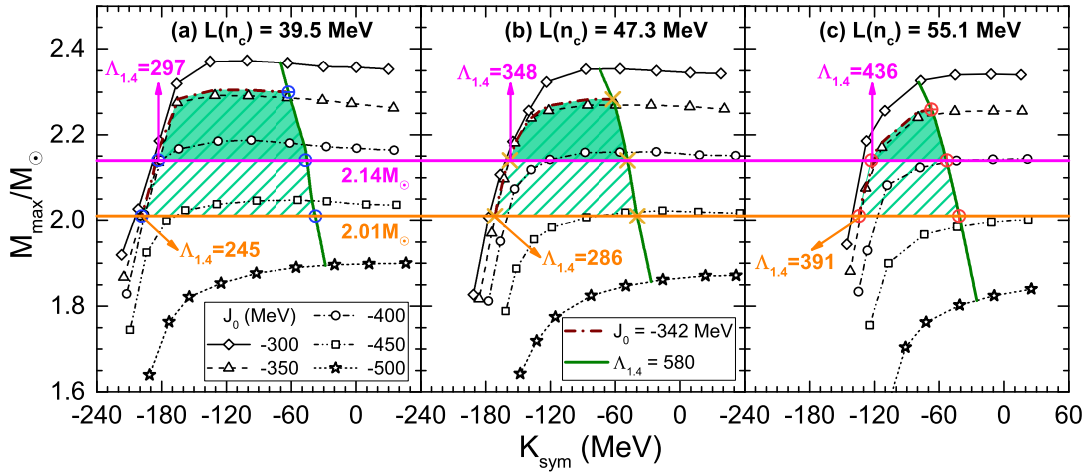}
\caption{NS maximum mass $M_{\rm max}$ vs $K_{\rm sym}$
within the eSHF model in a series of extended Skyrme interactions with
$J_0$ and $K_{\rm sym}$ fixed at various values
for $L(n_c) = 39.5$~MeV~(a), $47.3$~MeV~(b) and $55.1$~MeV~(c), respectively.
The shadowed regions indicate the allowed parameter space.
See the text for details.}
\label{fig1_Mmax-Ksym}
\end{figure*}

\subsection{Fitting strategy for model parameters}
In the eSHF model, there are totally $14$ model parameters, i.e.,
$n_0$, $E_{0}(n_0)$, $K_0$, $J_0$,
$E_{\rm sym}(n_r)$, $L(n_r)$, $K_{\rm sym}(n_r)$, $m_{s,0}^{\ast}$,
$m_{v,0}^{\ast}$, $G_S$, $G_V$, $G_{SV}$, $G_0'$, and $W_0$.
Following the same fitting strategy for model parameters as in Ref.~\citep{ZhouY19}, we first fix
$E_{\rm sym}(n_c) = 26.65$~MeV and $L(n_c) = 47.3$~MeV
at the subsaturation density $n_c=0.11n_0/0.16$
according to the precise constraint $E_{\rm sym}(n_c) = 26.65\pm 0.2$ MeV~\citep{Zha13}
by analyzing the binding energy difference of heavy isotope pairs and
$L(n_c) = 47.3 \pm 7.8$ MeV~\citep{Zha14} extracted from the electric dipole
polarizability of $^{208}$Pb.
In addition,
the higher-order parameters $J_0$ and $K_{\rm sym}$ generally have small influence
on the properties of finite nuclei but are crucial for the high-density nuclear
matter EOS and the NS properties.
To explore the effects of $J_0$ and $K_{\rm sym}$,
we thus fix them at various values
but with the other $10$ parameters being obtained by fitting
the data of finite nuclei
by minimizing the weighted sum of
the squared deviations of the theoretical predictions
from the experimental data, i.e.,
\begin{equation}
\chi^2(\bm{p})=\sum_{i=1}^{N}\left(\frac{{\cal O}_{i}^{\rm th}({\bm p})
-{\cal O}_{i}^{\rm exp}}{\triangle {\cal O}_i}\right)^2,
\end{equation}
where the $\bm{p}=(p_1 ,...,p_z) $ denote
the $z$ dimensional model space,
${\cal O}_i^{({\rm th})}$ and ${\cal O}_i^{({\rm exp})}$ are
the theoretical predictions and the corresponding experimental data,
respectively, and $\Delta {\cal O}_i$ is the adopted error
for balancing the relative weights of different
types of observables (see, e.g., Ref.~\citep{Zha16}).
The $30$ data of finite nuclei used in this work are listed in
Table~\ref{expdata}.
As for $\Delta {\cal O}_i$, we use $1.0$ MeV and $0.01$ fm for
the $E_{\rm{B}}$ and $r_{\rm{c}}$, respectively,
and for the $E_{\rm{GMR}}$ we use the experimental error
multiplied by $3.5$ to also consider the effect of the experimental error,
while for the $\epsilon_{\rm{ls}}^A$ a $10\%$ relative error is employed.
Considering the relatively larger uncertainty for $L(n_c) = 47.3 \pm 7.8$ MeV~\citep{Zha14},
we also investigate the cases with $L(n_c) = 39.5$~MeV and $55.1$~MeV.

\section{Results and discussions}
Using the fitting strategy described before,
for $E_{\rm sym}(n_c) = 26.65$ MeV and $L(n_c) = 39.5$~MeV, $47.3$ MeV, and $55.1$ MeV,
we construct a series of extended Skyrme parameter sets
with fixed $J_0$ in the large range of ($-500,-300$)~MeV and
$K_{\rm sym}$ in ($-220,60$) MeV.
As found in Ref.~\citep{ZhouY19},
in order to be consistent with the constraint on the pressure of SNM in the density region of
about $2n_0 \sim 5n_0$ from the flow data in heavy-ion
collisions~\citep{Dan02}, the $J_0$ must be less than $-342$ MeV, i.e.,
the upper limit of $J_0$ is $J^{\rm up}_{0} = -342$ MeV, independent of the values of
$L(n_c)$ and $K_{\rm sym}$.
Therefore, the flow data put strong constraint on the EOS of SNM at suprasaturation densities
and can significantly limit the maximum mass of NSs~\citep{ZhouY19}.

Shown in Fig.~\ref{fig1_Mmax-Ksym} is
the NS maximum mass $M_{\rm max}$ vs $K_{\rm sym}$
using various extended Skyrme parameter sets.
It is seen that for each $L(n_c)$ with a fixed $J_0$,
the $M_{\rm max}$ becomes insensitive to $K_{\rm sym}$ when the latter is larger than
a critical value $K^{\rm crit}_{\rm sym}$. For $L(n_c) = 39.5$~MeV, $47.3$ MeV, and $55.1$ MeV,
the value of $K^{\rm crit}_{\rm sym}$ is roughly $-130$ MeV, $-100$ MeV, and $-70$ MeV, respectively.
These results imply that the $E_{\rm sym}(n)$ has little influence on the $M_{\rm max}$
when the $K_{\rm sym}$ is large enough.
This can be understood from the fact that for the stiff high-density $E_{\rm sym}(n)$
with large $K_{\rm sym}$, the NS matter becomes almost isospin
symmetric at high densities and the $M_{\rm max}$ hence
essentially depends on the high-density EOS of SNM,
which is mainly controlled by the $J_0$.

On the other hand,
it is very interesting to see that for a fixed $J_0$, the $M_{\rm max}$ decreases drastically
as the $K_{\rm sym}$ decreases when the $K_{\rm sym}$ is less than $K^{\rm crit}_{\rm sym}$.
This means that the observed heaviest NS mass can rule out too soft high-density $E_{\rm sym}(n)$
with small $K_{\rm sym}$ values.
From Fig.~\ref{fig1_Mmax-Ksym}, one sees that for a fixed $K_{\rm sym}$, the $M_{\rm max}$ generally
increases with $J_0$.
Consequently, the extended Skyrme parameter sets with $J_0 = J^{\rm up}_{0} = -342$ MeV
generally predict the largest $M_{\rm max}$ in the eSHF model.
For $L(n_c) = (39.5, 47.3, 55.1)$~MeV, we obtain
the largest $M_{\rm max}$ in the eSHF model as $(2.30, 2.28, 2.26)M_\odot$.
Furthermore,
we find for $L(n_c) = (39.5, 47.3, 55.1)$~MeV,
using the recently discovered heaviest NS with mass $2.14M_\odot$ sets
a lower limit of $K_{\rm sym}$, namely,
$K^{\rm low}_{\rm sym} = (-183, -157, -123)$~MeV,
while using a NS maximum mass $2.01M_\odot$
gives $K^{\rm low}_{\rm sym} = (-198, -171, -134)$~MeV.
Therefore, the existence of heavier NSs requires a stiffer lower bound of
the high-density $E_{\rm sym}(n)$ with larger $K^{\rm low}_{\rm sym}$.

In addition, we note that
for each $L(n_c)$, the $\Lambda_{1.4}$ monotonically increases with $K_{\rm sym}$ ($J_0$)
for a fixed $J_0$ ($K_{\rm sym}$)
but the sensitivity on $K_{\rm sym}$  is much stronger than that on $J_0$~\citep{ZhouY19}.
Therefore, the existence of the lower limit for $K_{\rm sym}$ (i.e., $K^{\rm low}_{\rm sym}$)
leads to a lower bound of $\Lambda_{1.4}$, namely,
$\Lambda^{\rm low}_{1.4} = (297, 348, 436)$
for $L(n_c) = (39.5, 47.3, 55.1)$~MeV
based on the so far measured heaviest NS mass $2.14M_\odot$.
On the other hand, the lower bound of $\Lambda_{1.4}$ is found to be
$\Lambda^{\rm low}_{1.4} = (245, 286, 391)$ for $L(n_c) = (39.5, 47.3, 55.1)$~MeV
by using a NS maximum mass $2.01M_\odot$.
These results show that the lower bound of $\Lambda_{1.4}$ changes from $\Lambda^{\rm low}_{1.4} = 286^{+105}_{-41}$ to
$\Lambda^{\rm low}_{1.4} = 348^{+88}_{-51}$
when the measured largest NS mass varies from $2.01M_\odot$ to $2.14M_\odot$.
Therefore,
the recently discovered heaviest NS, i.e., PSR J0740+6620~\citep{Cro19Mmax},
puts a much stronger limit on $\Lambda^{\rm low}_{1.4}$,
i.e., $\Lambda_{1.4} \ge 348^{+88}_{-51}$.
The quite large lower bound
of $\Lambda_{1.4} \ge 348^{+88}_{-51}$ combined with the upper limit $\Lambda_{1.4} \le 580$~\citep{Abb18NSMerger}
from the GW signal GW170817 leads to a stringent constraint on the $\Lambda_{1.4}$,
i.e., $348^{+88}_{-51} \le \Lambda_{1.4} \le 580$.
This will have important implications on the structure
properties of NSs and the NS-involved GW detection in future.

Since the $\Lambda_{1.4}$ rapidly increases with $K_{\rm sym}$,
the upper limit $\Lambda_{1.4} \le 580$ from the GW signal GW170817~\citep{Abb18NSMerger}
can set upper limits on $K_{\rm sym}$
for various values of $J_0$ as shown in Fig.~\ref{fig1_Mmax-Ksym}.
According to the allowed parameter space shown
in Fig.~\ref{fig1_Mmax-Ksym}, the recently discovered heaviest NS
with mass $2.14M_\odot$ sets a upper limit of $K_{\rm sym}$, namely,
$K^{\rm up}_{\rm sym} = (-46, -48, -53)$~MeV for $L(n_c) = (39.5, 47.3, 55.1)$~MeV.
We note that using a NS maximum mass $2.01M_\odot$
gives $K^{\rm up}_{\rm sym} = (-37, -39, -42)$~MeV for $L(n_c) = (39.5, 47.3, 55.1)$~MeV.
The existence of the upper and lower limits of $K_{\rm sym}$ can
rule out too stiff and too soft high-density $E_{\rm sym}(n)$ and
thus put strong constraints on the high-density behaviors of $E_{\rm sym}(n)$.

\begin{figure}[tbp]
\centering
\includegraphics[width=0.9\linewidth]{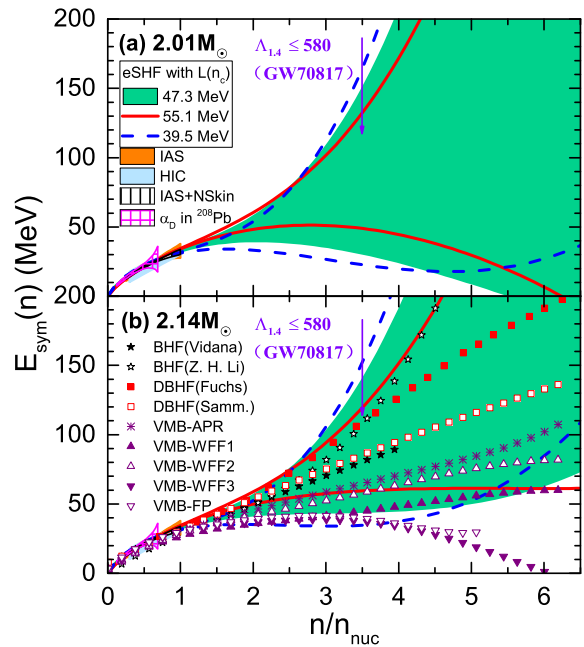}
\caption{Density dependence of the symmetry energy by assuming $M_{\rm max} = 2.01M_\odot$ (a) and $2.14M_\odot$ (b).
See the text for details.}
\label{fig2_Esym}
\end{figure}

Figure~\ref{fig2_Esym} shows the density dependence of the symmetry energy
according to the allowed parameter space for $J_0$ and $K_{\rm sym}$ with $L(n_c) = (39.5, 47.3, 55.1)$~MeV
as shown in Fig.~\ref{fig1_Mmax-Ksym}.
Fig.~\ref{fig2_Esym}~(a) is obtained by using $2.01M_\odot$ as the NS maximum mass
while Fig.~\ref{fig2_Esym}~(b) is by using $2.14M_\odot$.
Also included in Fig.~\ref{fig2_Esym} are
the constraints on the $E_{\rm sym}(n)$ at subsaturation densities from
midperipheral heavy-ion collisions of Sn isotopes~\citep{Tsa09},
the isobaric analog states (IAS) and combining the neutron skin
data (IAS + NSkin)~\citep{Dan14}, and the electric dipole
polarizability ($\alpha_D$) in $^{208}$Pb \citep{Zha15}.
For comparison, we further include in Fig.~\ref{fig2_Esym}~(b) the results from some microscopic many-body
approaches, namely, the non-relativistic Brueckner-Hartree-Fock (BHF) approach~\citep{Vid09,Li08},
the relativistic Dirac-Brueckner-Hartree-Fock (DBHF) approach~\citep{Kla06,Sam10}, and
the variational many-body (VMB) approach~\citep{APR98,Fri81,Wir88}.
It is seen from Fig.~\ref{fig2_Esym} that the $E_{\rm sym}(n)$ with
various values of $L(n_c)$, $J_0$ and $K_{\rm sym}$ in the allowed parameter space
are all in good agreement with
the experimental constraints at subsaturation densities
but exhibit very different high-density behaviors.

From Fig.~\ref{fig2_Esym}~(a), one can see that in the case with a NS maximum mass $2.01M_\odot$,
the lower bound of the $E_{\rm sym}(n)$ becomes negative when the density is larger than
$n/n_{\rm nuc} \approx (5.6, 6.3)$ for $L(n_c) = (47.3, 55.1)$~MeV
(Here $n_{\rm nuc} = 0.16$~fm$^{-3}$ represents nuclear normal density).
We note the corresponding central density $n_{\rm cen}$ of the NS with mass $2.01M_\odot$ is
$n_{\rm cen}/n_{\rm nuc} \approx (6.4, 7.4)$ for $L(n_c) = (47.3, 55.1)$~MeV,
indicating that the lower bound of the $E_{\rm sym}(n)$ already becomes negative at suprasaturation
densities inside the NS,
and therefore the corresponding $E_{\rm sym}(n)$ is supersoft,
which can cause the appearance of a PNM core in the NS
(Note: the higher-order symmetry energies, e.g., the fourth-order symmetry energy~\citep{Cai12},
may affect the proton fraction in NS matter, and especially in the case of the supersoft symmetry energy,
they may obviously change the disappearance density of the proton fraction in NSs~\citep{Zha01}).
Our results thus demonstrate that
the supersoft high-density $E_{\rm sym}(n)$ can support a NS with mass $2.01M_\odot$,
and at the same time it can describe very successfully
the data of finite nuclei and agree well with the constraint from the flow data in heavy-ion collisions.

In the case with a NS maximum mass $2.14M_\odot$, on the other hand,
it is very interesting to see from Fig.~\ref{fig2_Esym}~(b)
that the $E_{\rm sym}(n)$ is always positive and the supersoft $E_{\rm sym}(n)$ is clearly ruled out.
This means that
the eSHF model with a supersoft $E_{\rm sym}(n)$ cannot simultaneously describe
the data of finite nuclei, the constraint on SNM EOS from flow data in heavy-ion
collisions, and the NSs with mass $2.14M_\odot$.
Our results therefore exclude the possibility for the appearance of a PNM core in NSs.
Furthermore,
while our results are consistent with most of
the microscopic many-body calculations shown in Fig.~\ref{fig2_Esym}~(b),
they indeed rule out the VMB calculations with interactions WFF1 (i.e., AV14 plus UVII),
WFF3 (i.e., UV14 plus TNI)~\citep{Wir88} and FP (i.e., $v_{14}$ + TNI)~\citep{Fri81}.
Our present results
also rule out many non-relativistic Skyrme and Gogny effective interactions that predict
negative symmetry energy at suprasaturation densities (See, e.g., Refs.~\citep{Sto03,ChenLW17}).
It is interesting to note that
our results seem to support the relativistic mean-field description
of nuclear matter, which generally cannot predict
negative $E_{\rm sym}(n)$ at high densities
due to the specific construction of meson exchanges~\citep{ChenLW07,Dut14,ChenLW17}.
Our present results on the constraints of high-density $E_{\rm sym}(n)$
may also have important implications on the poorly known
effective three-body forces, short-range tensor forces and
short-range nucleon-nucleon correlations~\citep{XuC10,Cai18}.

Finally, we would like to point out that
including new degrees of freedom such as
hyperons~\citep{Vid11,Lon15}, antikaon condensates~\citep{Gup13,Cha14},
and quark matter~\citep{Bom16,Alf17,Dex18} that could be present in
the interior of NSs but neglected in the present work,
usually softens the NS matter EOS,
and in this case a stiffer high-density
$E_{\rm sym}(n)$ would be necessary to obtain a NS with mass $2.14M_{\odot}$.
Therefore, including the new degrees of freedom in NSs
is also expected to rule out the supersoft high-density $E_{\rm sym}(n)$.

\section{Conclusion}
Within the theoretical framework of the eSHF model,
we have demonstrated that a supersoft high-density symmetry energy cannot simultaneously
describe the data of finite nuclei, the equation of state of symmetric nuclear
matter at suprasaturation densities constrained from the flow data in heavy-ion
collisions, and the maximum neutron star mass of $2.14M_\odot$,
although it is still allowed
if the maximum neutron star mass is $2.01M_\odot$.
Therefore, the very recent discovery of PSR J0740+6620 rules out
the supersoft high-density symmetry energy,
which means it is unlikely to have a pure neutron matter core in neutron stars.
Furthermore,
we have found that the stiffer lower limit of the high-density symmetry energy based on the existence of
$2.14M_\odot$ neutron stars leads to a quite large lower limit for $\Lambda_{1.4}$, i.e.,
$\Lambda_{1.4} \ge 348^{+88}_{-51}$, which is expected to have important implications on
the future multimessenger observations of neutron-star-involved GW events.

\acknowledgments
The authors thank Tanja Hinderer, Bao-An Li and Zhen Zhang
for useful discussions.
This work was supported in part by the National Natural Science
Foundation of China under Grant No. 11625521, the Major State Basic Research
Development Program (973 Program) in China under Contract No.
2015CB856904, the Program for Professor of Special Appointment (Eastern
Scholar) at Shanghai Institutions of Higher Learning, Key Laboratory
for Particle Physics, Astrophysics and Cosmology, Ministry of
Education, China, and the Science and Technology Commission of
Shanghai Municipality (11DZ2260700).




\end{document}